\documentclass[a4,11pt]{reportN}
\usepackage[english]{babel}    

\usepackage[utf8]{inputenc}   

\usepackage{graphics}
\usepackage{subfigure}
\usepackage{graphicx}
\usepackage{epsfig}
\usepackage[centertags]{amsmath}
\usepackage{graphicx,indentfirst,amsmath,amsfonts,amssymb,amsthm,newlfont}
\usepackage{longtable}
\usepackage{cite}
\usepackage[usenames,dvipsnames,svgnames,table]{xcolor}

\usepackage{multicol}
\usepackage{verbatim}

\begin{document}

\title{Synchronization on the accuracy of chaotic oscillators simulations}
\author
{ \Large{Gabriel Hugo Álvares Silva}\thanks{gabrielhugo0701@gmail.com} \\
\Large{Igor Carlini Silva}\thanks{igorufsj@yahoo.com.br}\\
\Large{Wilson Rocha Lacerda Junior}\thanks{wilsonrljr@outlook.com}\\
\Large{Samir Angelo Milani Martins}\thanks {martins@ufsj.edu.br} \\
\Large{Márcio Falcão Santos Barroso}\thanks {barroso@ufsj.edu.br}\\
\Large{Erivelton Geraldo Nepomuceno}\thanks {nepomuceno@ufsj.edu.br}\\
{\small Control and Modeling Group, UFSJ, São João del-Rei, MG, Brazil\\
  Department of Electrical Engineering, UFSJ, São João del-Rei, MG, Brazil} }

\criartitulo


\begin{abstract}
{\bf Abstract}. Numerical problems are considered on general synchronization of chaotic oscillators, through the evaluation of the Lower Bound Error index on two case studies: a Lorenz system unidirectionally coupled to a Duffing system and a Duffing system unidirectionally coupled to a Rossler system. It was possible to observe, in each case, that the behavior of the slave's LBE curve tends to follow the behavior of the master's as the value of the coupling constant is increased up to a certain value, and thus, that synchronization can affect numerical calculations.



\end{abstract}


\section{Introduction}

Since the work of Lorenz in 1963 \cite{lorenz1963deterministic}, chaos synchronization has been extensively studied by several researchers. Synchronization of chaos is often understood as a regime in which two coupled chaotic systems exhibit identical, but still chaotic, oscillations \cite{pecora1990}. Chaos synchronization has been applied in electrical \cite{XIAO2009}, biological, chemical, and secure communication \cite{naderi2016exponential} problems.

In this context, it is well known that countless researchers identifies a chaotic system behaviour \cite{ott2002chaos} by analyzing numerical solutions, obtained using popular software, in which the reliability of results is not carefully verified \cite{lozi2013can}. Nepomuceno \cite{nepomuceno2014convergence} shows that a simple sequence of iterations of discrete logistic model may generate a steady state result that converges to the wrong answer. It is worth noting that investigation of propagation error is not a recent issue \cite{hammel1987numerical}. In fact, there are many works based on deterministic or stochastic tools that provide some confidence in simulation of recursive functions.

Based on the fact that although interval extensions are mathematically equivalent, they may generate different computer simulation outcomes, Nepomuceno and Martins \cite{nepomuceno2016lbe} introduced an approach to evaluate a lower bound error in recursive functions. The Lower Bound Error (LBE) index may conduce the understanding of the solutions generated by nonlinear dynamical systems. Nepomuceno and Mendes \cite{nepomuceno2017analysis} show the existence of multiple pseudo-orbits for nonlinear dynamics systems when discretization schemes are used, when the step-size and the initial conditions are kept unchanged. 

Although there are many studies regarding synchronization of chaotic oscillators and regarding numerical problems, as far as we know, no study deals with numerical problems in the synchronization of chaotic systems. This is the main scope of this paper. We developed two cases studies in which we coupled a Lorenz system to a Duffing system and a Duffing system to a Rossler system, in order to evaluate the effects of general synchronization on the reckoning of Lower Bound Error.

This paper is laid out as follows: In Section 2, the preliminary concepts are briefly reviewed. The proposed method based on LBE is presented in Section 3. The results as well as the discussion are presented in Section 4, while concluding remarks and perspectives for future research are shown in Section 5.

\section{Preliminary Concepts}

\subsection{Generalized synchronization}

The simplest concept of synchronization between chaotic oscillators, called complete synchronization, occurs when the distance between the state variables of two dynamical systems converges to zero while evolving in time. Let $\dot x=F(x)$ and $\dot y=G(y)$ be the representations of two chaotic systems, where $x$ is a \textit{n}-dimensional state vector and $y$ is a \textit{m}-dimensional state vector. F and G are vector fields, $F:R^n\rightarrow R^n$, and $G:R^m\rightarrow R^m$. Those systems are said to be completely synchronized if $lim_{t \rightarrow \infty}||x(t)-y(t)||=0$.

In \cite{rulkov1995}, Rulkov et al presented a generalization of the definition of complete synchronization, where the state variables of the systems considered do not have to be identical. For this case, it is sufficient if they present a functional relation. Let $\dot x = F(x)$ and $\dot y=G(y, h_{\mu}(x))$ be two unidirectionally coupled systems, where $x$ is the \textit{n}-dimensional state vector of the driver and $y$ is the \textit{m}-dimensional state vector of the response. F and G are vector fields, $F:R^n\rightarrow R^n$, and $G:R^m\rightarrow R^m$. The vector field $h_{\mu}(x):R^m\rightarrow R^m$ rule the couple between response and driver. When the parameter $\mu=0$, both systems are chaotic, since there is no relation between their evolution. When $\mu~=0$, the systems are considered generally synchronized if exists a transformation $\psi:x\rightarrow y$ which is able to map asymptotically the trajectories of the driver attractor into the ones of the response \cite{boccaletti2002}.

\subsection{Lower Bound Error}

Before introducing LBE's concept, we need to present the definitions of orbits and pseudo-orbits. An orbit is a sequence of values of a map, represented by {\footnotesize{${x_n}=[x_0,x_1,x_2,...,x_n]$}}. A pseudo-orbit is an approximation of an orbit and is expressed as {\footnotesize{${\hat x_{i,n}}=[\hat x_{i,0},\hat x_{i,1},\hat x_{i,2},...,\hat x_{i,n}]$}}.

The Lower Bound Error (LBE) is a method presented by Nepomuceno and Martins in \cite{nepomuceno2016lbe}, which aims to evaluate the error propagation due to round off in digital computers. The procedure to calculate the LBE is based on the comparison of two pseudo-orbits produced from two mathematical equivalent models, but different from the point of view of floating point representation. Therefore, LBE's mathematical representation in given by $2\delta_{\alpha,n}=|\hat x_{a,n} - \hat x_{b,n}|$, where $\delta_{\alpha,n}$ represents the lower bound error between two pseudo-orbits $\hat x_{a,n}$ e $\hat x_{b,n}$.




\section{Methodology}

In order to verify the influence of synchronization on the reckoning of Lower Bound Error, two case studies were considered. In the first one, a Lorenz system (slave) was unidirectionally coupled to a Duffing system (master). In the second case, the same procedure was followed for a Duffing system (slave) and a Rossler system (master).

\subsection{Duffing - Lorenz}

The Duffing and Lorenz systems are given by Equations (\ref{eq:duffing1}) and (\ref{eq:lorenz1}), respectively: 

\begin{multicols}{2}
\noindent \begin{eqnarray} \dot x_1 & = & x_2 \nonumber \label{eq:duffing1}\\ \dot x_2 & = & x_1-x_1^3-\delta x_2+\gamma cos(\omega t),\label{eq:duffing2}\end{eqnarray}
\noindent \begin{eqnarray} \dot y_1 & = & -10y_1+10y_2 \nonumber \label{eq:lorenz1}\\ \dot y_2 & = & 28y_1-y_2-y_1y_3+Kx_1  \label{eq:lorenz2}\\\dot y_3 & = & y_1y_2-\frac{8}{3}y_3. \nonumber \label{eq:lorenz3}\end{eqnarray}
\end{multicols}




To couple the systems, we added the term $Kx_1$ on the equation $\dot y_2$ of the Lorenz system, where $K$ is called coupling constant and determines how strong the synchronization between the oscillators is \cite{pyragas1996}. The state variable $x_1$ was chosen arbitrarily and could be $x_2$ as well. The initial conditions used for the Duffing system were $x_1(0)=3$ and $x_2(0)=4$ and for the Lorenz system, $y_1(0)=y_2(0)=y_3(0)=1$.

During the experiments, we increased the value of K from zero, when the systems are completely unsynchronized, up to a value where general synchronization is observed. In order to verify if the systems are in fact synchronized, we considered the method presented in \cite{pyragas1996}. The auxiliary equation used is given by $\dot y'_2=28y'_1-y'_2-y'_1y'_3+Kx_1$. The initial conditions used to compute the auxiliary equation were $y_1(0)=y_2(0)=y_3(0)=5$. As Pyragas defines in \cite{pyragas1996}, the two systems can be considered to be strongly synchronized in a general manner if complete synchronization exists between $\dot y_2$ and $\dot y'_2$.


Finally, LBE was calculated for two pseudo-orbits of the response system for different values of K. The pseudo-orbits were obtained by a simple mathematical manipulation on $\dot y_1$. The equation obtained after the manipulation is given by $\dot y_1=10(y_2-y_1)$. Thereby, we calculated the LBE using the Equations given by $\dot y_2$ of the two pseudo-orbits.



\subsection{Rossler - Duffing}

We followed the same procedures described previously to study the behavior of a Duffing system being driven by a Rossler system. The equations of Rossler circuit are given by $\dot x_1=-x_2+x_3$, $\dot x_2 =x_1+0.2x_2$ and $\dot x_3=0.2+x+3(x_1-5.7)$.

The Duffing system is the same used in the previous case of study (Equation (\ref{eq:duffing1})), with the addition of the coupling term to $\dot x_1$, thus, it became $\dot x_1=x_2+Ky_1$. We used the same initial conditions as in Section 3.1. The auxiliary equation is $\dot x'_1=x'_2+Ky_1$ and, for this equation, the initial conditions are $x_1(0)=5$ and $x_2(0)=6$. The initial conditions used for the Rossler system are $x_1(0)=x_2(0)=x_3(0)=1$. To calculate LBE, we performed a mathematical manipulation on Equation (\ref{eq:duffing2}), thus it became $\dot x_2=x_1-x_1x_1x_1-\delta x_2+\gamma cos(\omega t)$. Finally, we calculated LBE between the Equations given by $\dot x_1$ of the two pseudo-orbits.



\section{Results}

\subsection{Duffing - Lorenz}

As we increased the value of K, we applied Pyragas' method to verify the synchronization between the oscillators. We observed that for values greater than 30, we could see a straight line on the phase portrait of $y$ and $y'$. Therefore, we chose $K=40$ to represent in this paper.

\begin{figure}[ht!]
\subfigure[Duffing and Lorenz dynamics for K = 0.\label{d1}]{
\includegraphics[width=7cm,height=3.15cm]{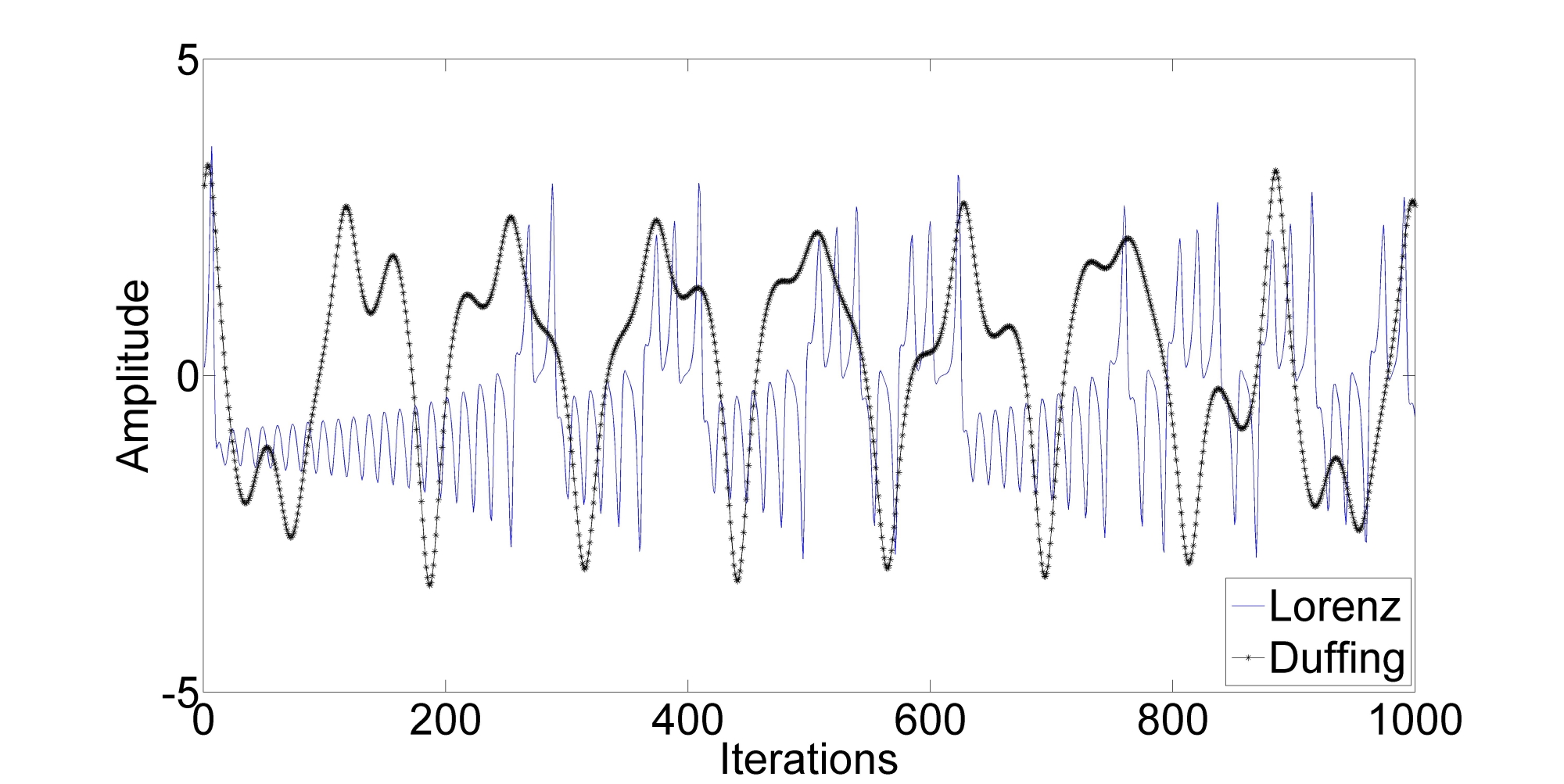}}
\subfigure[Duffing and Lorenz dynamics for K = 40.\label{d2}]{
\includegraphics[width=7cm,height=3.15cm]{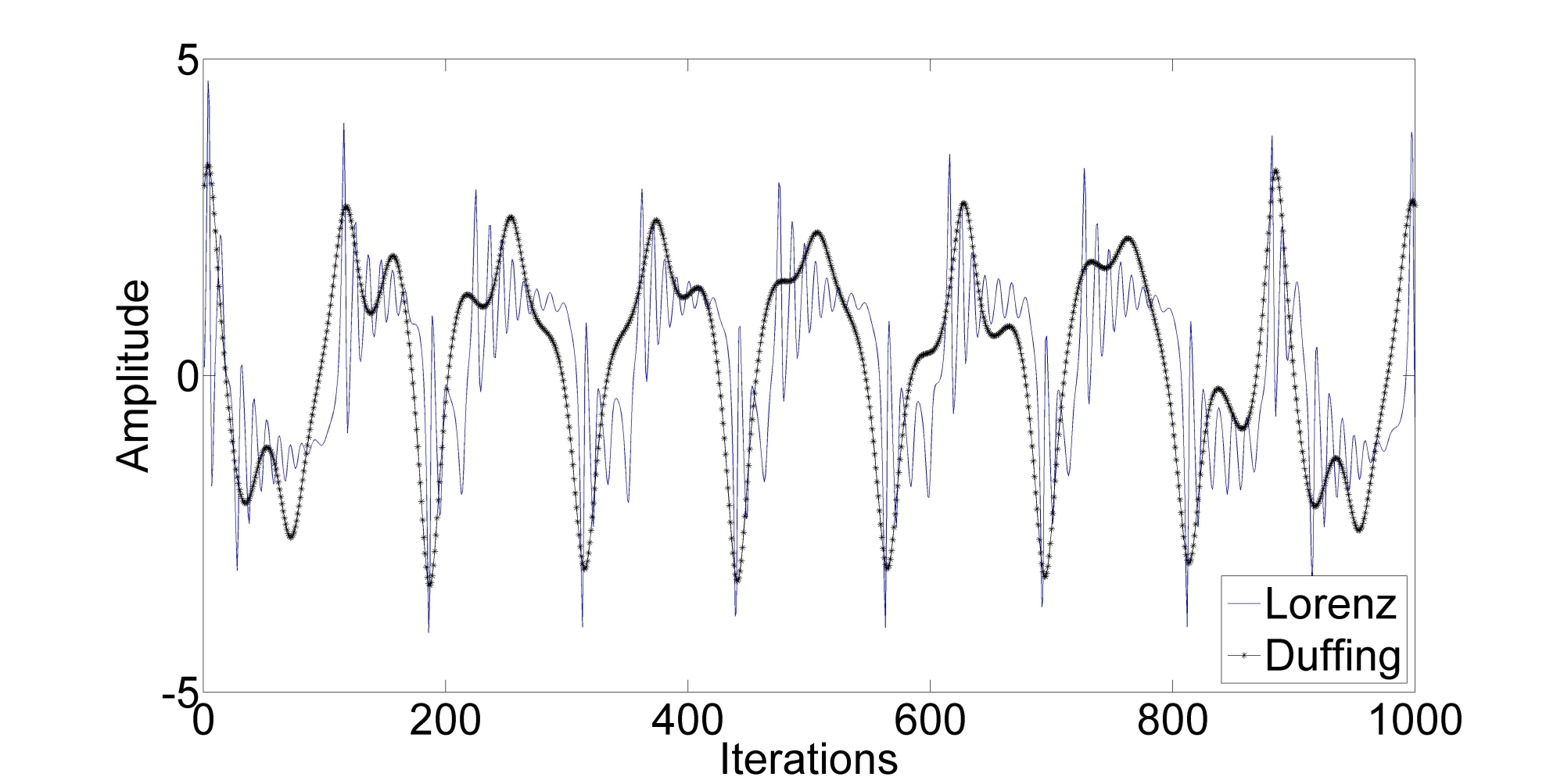}}
\caption{Dynamics of the systems when they are not synchronized (K=0) and when generalized synchronization exists (K=40).}
\end{figure}

\begin{figure}[ht!]
\subfigure[Phase portrait for K = 0.\label{yl1}]{
\includegraphics[width=6.9cm,height=3.15cm]{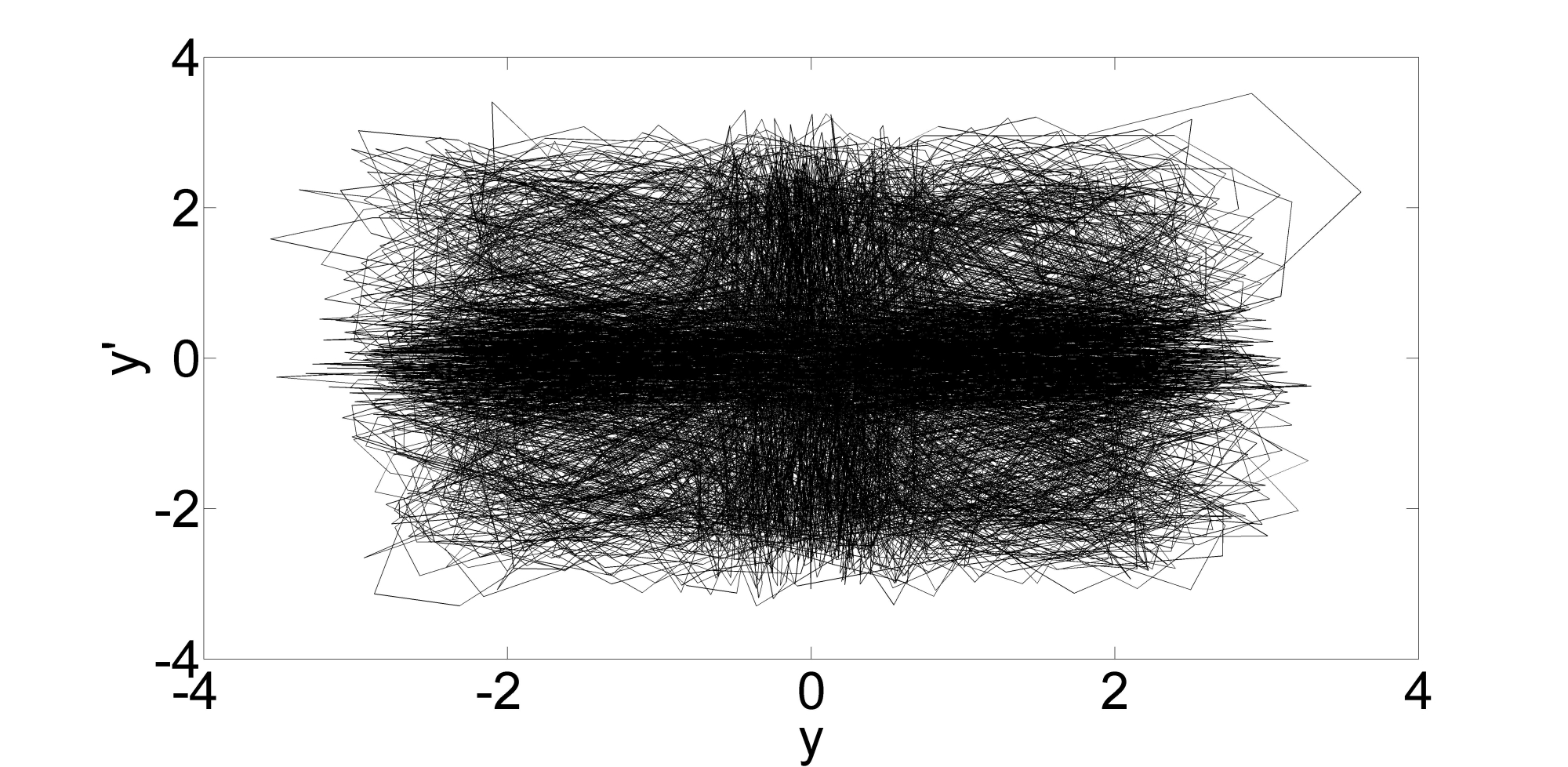}}
\subfigure[Phase portrait for K = 40.\label{yl2}]{
\includegraphics[width=6.9cm,height=3.15cm]{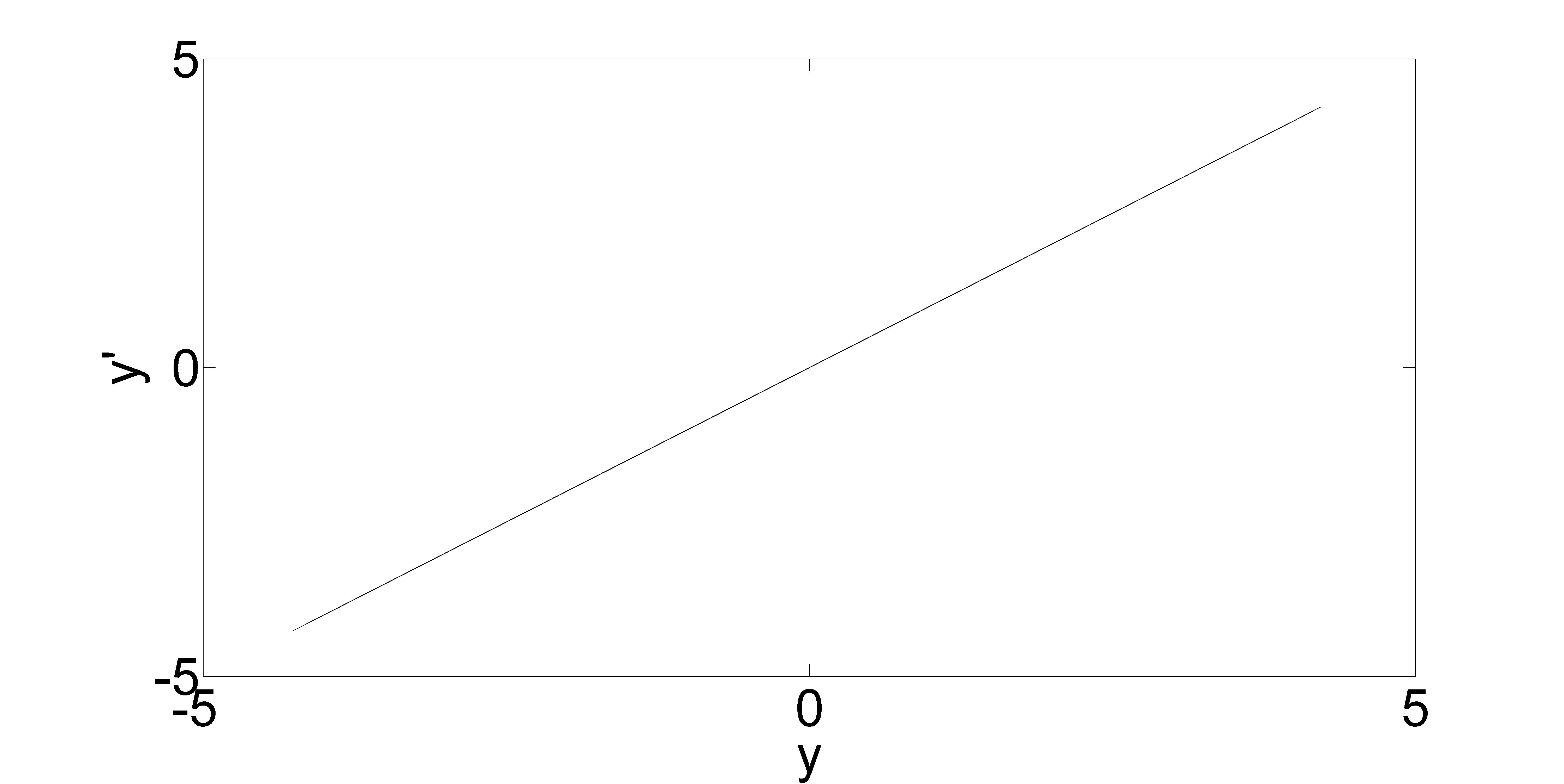}}
\caption{Synchronization between the systems using the auxiliary equation y'.}
\end{figure}

Figure \ref{d1} shows the dynamics of the systems when they are unsynchronized, that is, $K=0$. In figure \ref{d2}, we represent the dynamics of the systems when there is general synchronization between them for $k=40$. To validate the results, Figure \ref{yl1} shows the phase portrait for $K=0$, where $y$ and $y'$ are clearly unsynchronized. Figure \ref{yl2} represent the phase portrait of the systems for $K=40$, where a straight line is visible, which indicates complete synchronyzation between $y$ and $y'$, and, consequently, general synchronization between the Duffing and Lorenz systems.

\begin{figure}[ht!]
\centering
\includegraphics[width=12.85cm,height=4cm]{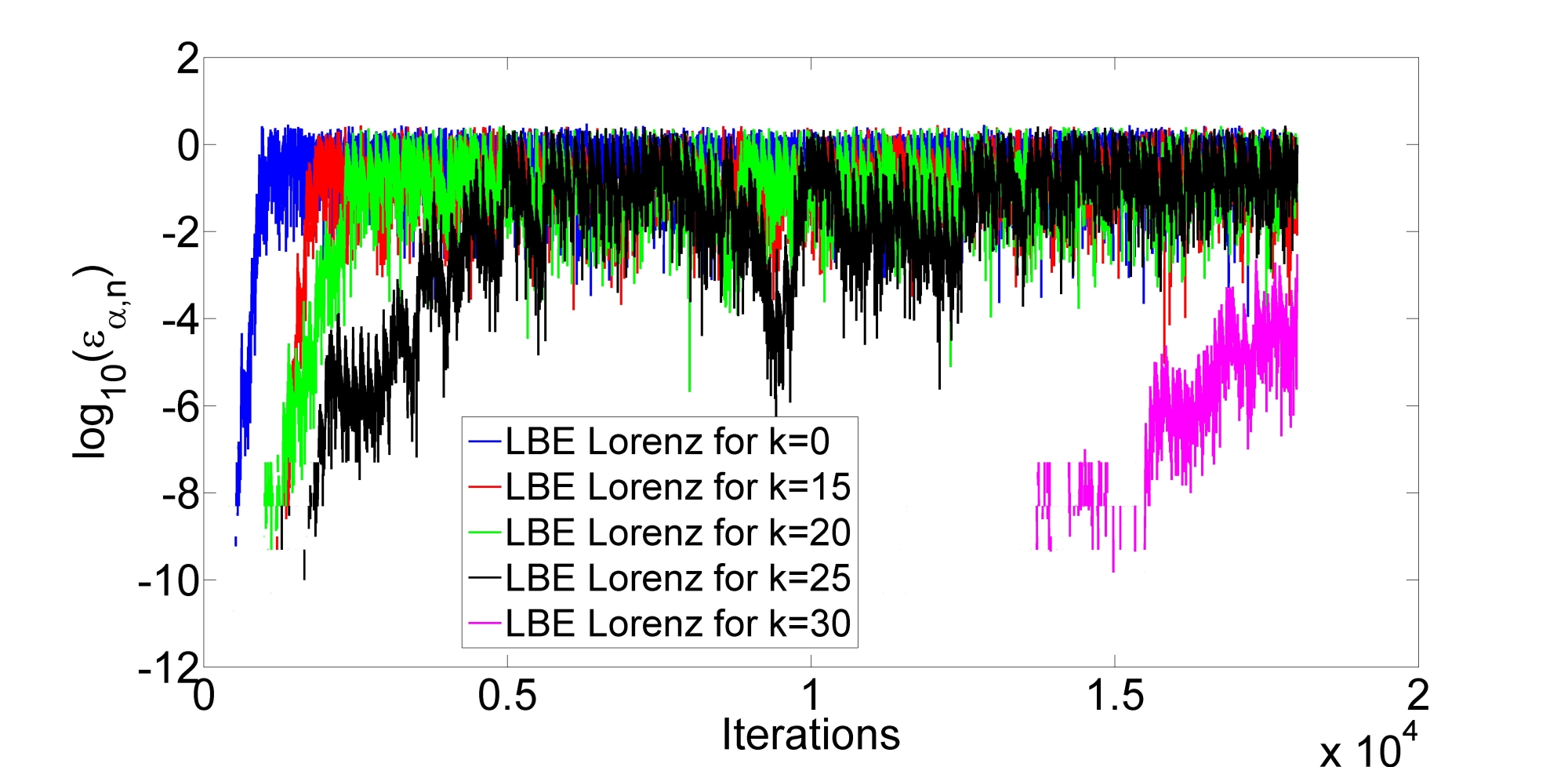}
\caption{Evolution of LBE according to value of K for Lorenz system.}
\label{lbeduff}
\end{figure}

The results of the computation of LBE between Equations $\dot x_2$ of the two pseudo-orbits for different values of K ranging from 0 to 30 are shown in Figure \ref{lbeduff}. As one can see, the LBE's curves tend to take more iterations to increase to higher values as K increase. For example, for $K=0$, it takes 1266 iterations for the LBE to go to -0.3 and for $K=25$, it takes 5100 iterations for the LBE to go to the same value. Apparently, the response's LBE curve is following the behavior of the master's. However, for $K=30$, Duffing's LBE curve will take about 5000 iterations to start increasing, as can be seen in Figure \ref{lbeduff1} for $K=0$ and, in Figure \ref{lbeduff}, for $K=30$, Lorenz's takes about 15000 iterations to start. This observation could mean that synchronization is also delaying the error propagation. It is also important to mention that, for values greater than 30, like 40, we were not able to represent LBE curve. Since we are using logarithmic notation to represent the results, this fact may imply that the Lower Bound Error goes to zero when the systems are strongly coupled.




\subsection{Rossler - Duffing}

\begin{figure}[ht!]
\subfigure[Rossler and Duffing dynamics for K = 400 \label{r12}]{
\includegraphics[width=7cm,height=3.15cm]{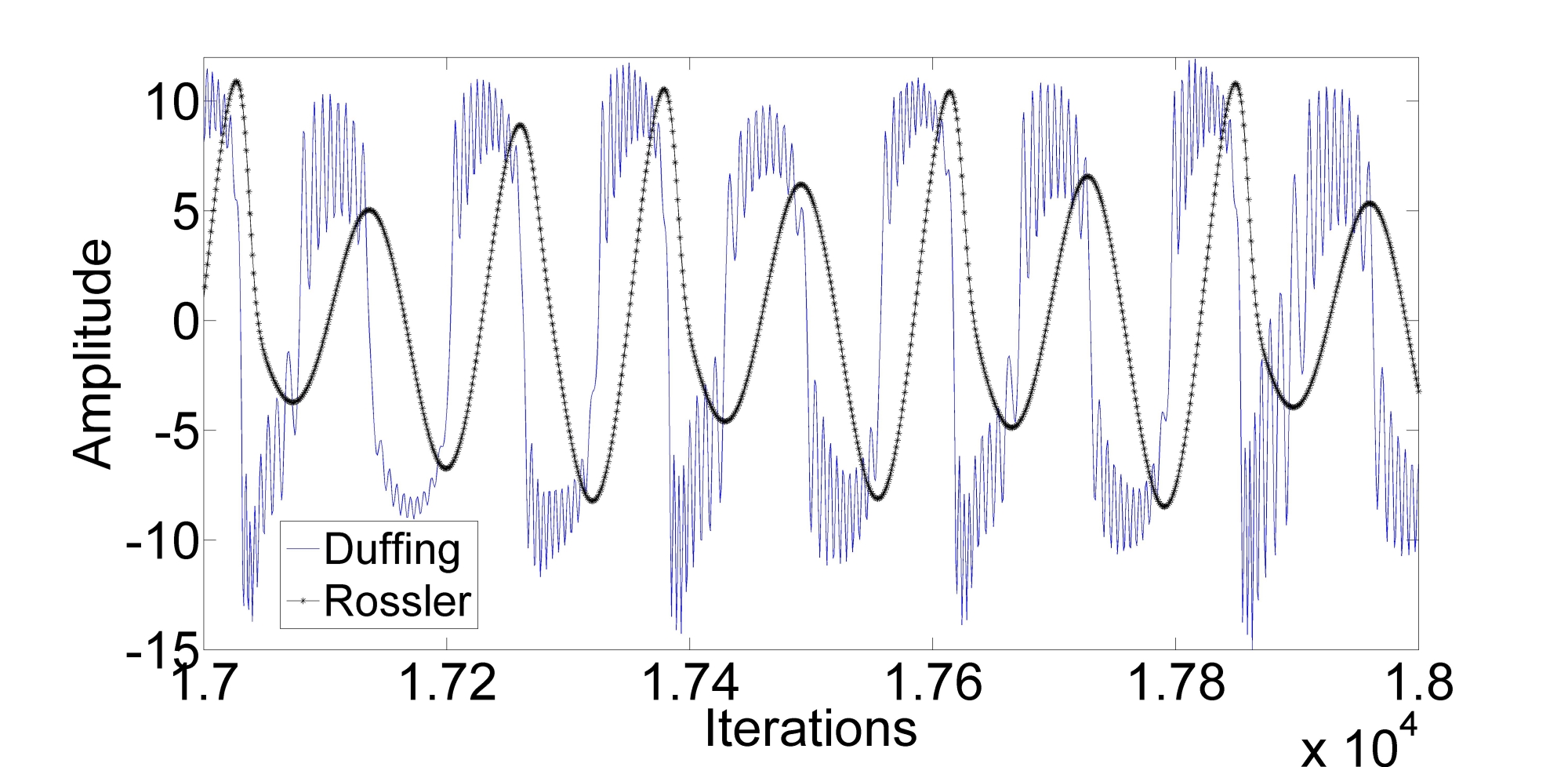}}
\subfigure[Phase portrait for K = 400.\label{r22}]{
\includegraphics[width=7cm,height=3.15cm]{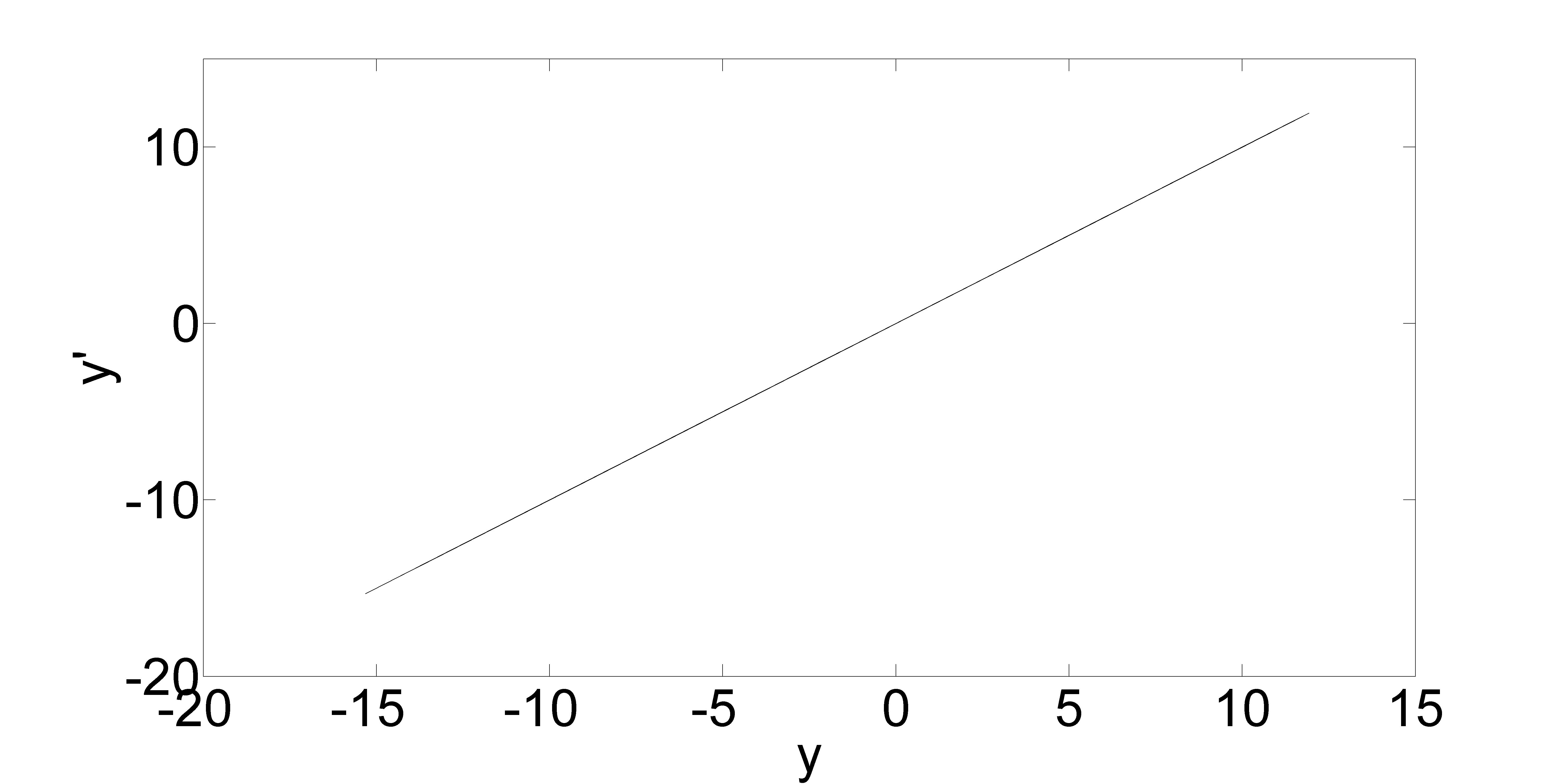}}
\caption{Dynamics of the systems and phase portrait between $y$ and $y'$.}
\label{ylduff}
\end{figure}

In order to analyze the behavior of Duffing system been driven by a Rossler, we followed the same procedure adopted in Section 4.1. In this case, we observed the occurrence of synchronization for values of K greater than 300. We chose 400 to represent. Figures \ref{r12} and \ref{r22} show the dynamics of the systems and the phase portrait for $k=400$.

\begin{figure}[ht!]
\centering
\includegraphics[width=12.85cm,height=4cm]{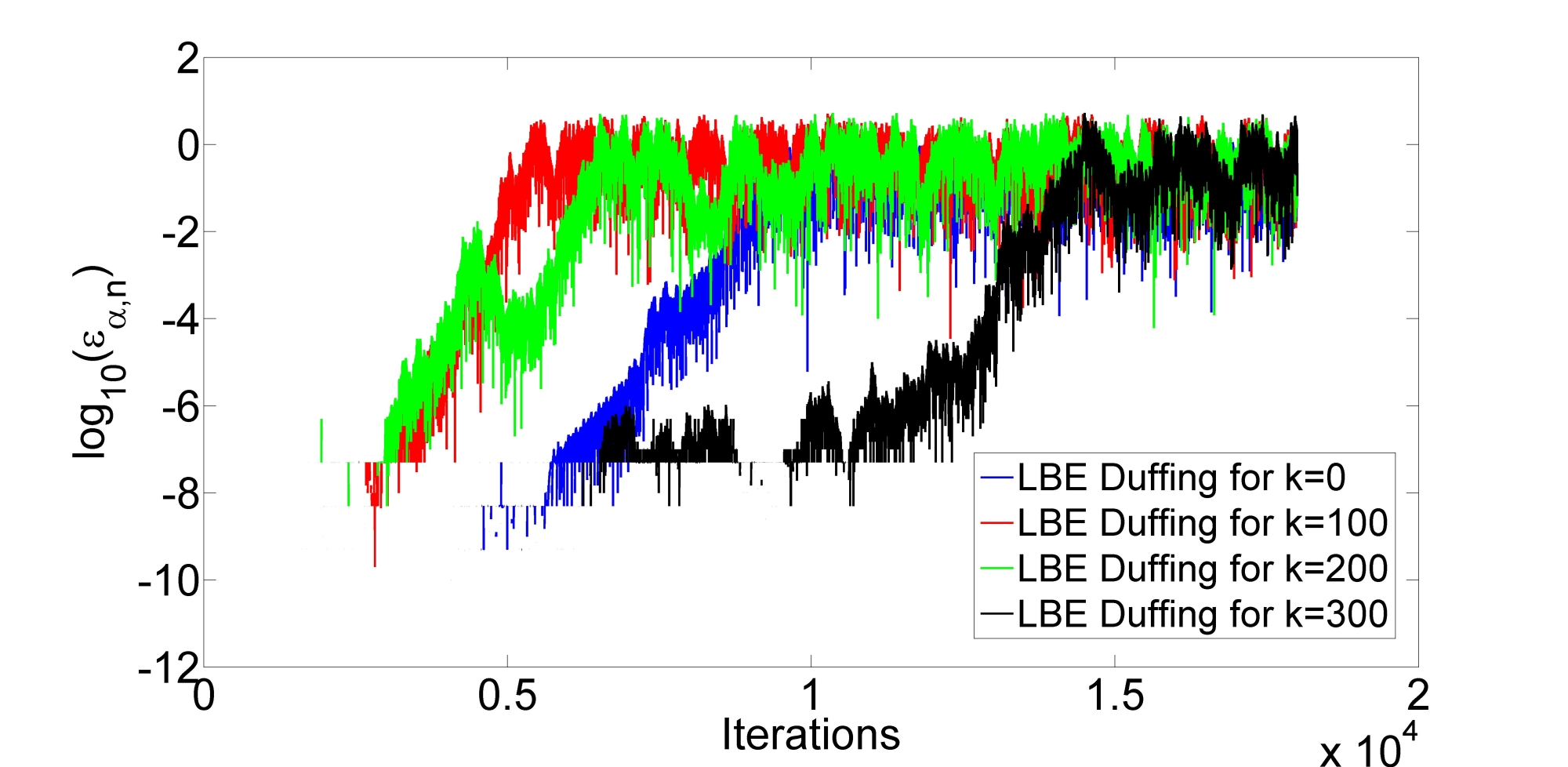}
\caption{Evolution of LBE according to value of K for Duffing system.}
\label{lbeduff1}
\end{figure}

The results of the computation of LBE between Equations $\dot x_2$ of the two pseudo-orbits for different values of k ranging from 0 to 300 are shown in Figure \ref{lbeduff1}. In this case, LBE's curve for $k=0$ takes 9698 iterations to go -0.3, while for $k=100$ and $k=200$, it takes about 3500 and 6500 iterations, respectively. Therefore, when the systems are weakly coupled, Duffing's curve will follow Rossler's. However, as K go as high as 300, for example, the curve will take about 14300 to get to -0.3. We were not able to represent LBE curve for values greater than 300. The same observations from Section 4.1 can be pointed in this case.





\section{Conclusions}

The results presented in Section 4 show that when a chaotic oscillator is synchronized to another, the LBE of two pseudo-orbits of the response system tend to follow the behavior of the driver until a certain value of K is reached  and, consequently, we demonstrated that synchronization can affect numerical calculations. We could observe that, in the two case studies investigated, after that value of K, the LBE index take more iterations to raise, what could mean that the LBE between the pseudo-orbits of the response system was decreasing. Therefore, synchronization can be further investigated to be used as a tool to reduce the error on numerical simulations. Also, in future works, other types of synchronization can be investigated, as complete, phase and lag synchronization.




		\section{Acknowledgement}
	The authors are thankful to the Brazilian agencies CNPq, FAPEMIG, and to UFSJ.
	





\end{document}